\documentstyle[prl,aps,epsf]{revtex}

\begin{document}


\twocolumn[\hsize\textwidth\columnwidth\hsize\csname@twocolumnfalse%
\endcsname

\title{Boundary Effects and the Order Parameter Symmetry of
High-T$_c$ Superconductors}
\author{Safi R. Bahcall}
\address{Department of Physics, University of California, 
Berkeley CA 94720}
\maketitle

\begin{abstract}

Apparently conflicting phase-sensitive measurements of the order
parameter symmetry in the high-T$_c$ cuprate superconductors may be
explained by regions near surfaces in which the order parameter
symmetry is different than in the bulk.  These surface states can lead
to interesting and testable effects.

\vspace*{.3in}

\end{abstract}


]

Phase-sensitive measurements on the high temperature superconductor
$\rm Y Ba_2^{} Cu_3^{} O_{7-\delta}^{}$ have yielded two potentially
conflicting sets of results for the symmetry of the superconducting
order parameter \cite{dvhreview}.  Measurements involving currents
flowing in the CuO$_2$ planes, such as the corner-junction SQUID
experiments \cite{wollman,brawner,mathai}, the corner-junction flux
modulation experiments \cite{wollmanfm,miller}, and the tricrystal
ring experiments \cite{tsuei} indicate an order
parameter with primarily $d_{x^2-y^2}$ symmetry under rotations in the
plane ($\Delta({\bf k})\sim \cos k_x-\cos k_y$).  The presence,
however, of Josephson tunneling perpendicular to the CuO$_2$ planes
between heavily-twinned YBCO and a conventional $s$-wave
superconductor \cite{dynes,iguchi,clarke} suggests an order parameter
with a significant $s$-wave component \cite{dontconsider,scalapino}.

A bulk order parameter of mixed $s$ and $d_{x^2-y^2}$ symmetry could
explain both sets of experiments.  An order parameter with this mixed
symmetry, for a material which is otherwise macroscopically symmetric
under $90^{\rm o}$ rotations (heavily twinned YBCO), requires either a
first order transition or two separate bulk phase transitions.  So
far, there has been no convincing evidence for either of these.  In
this paper, then, we assume that the order parameter in the bulk
superconductor transforms as one irreducible representation of the
rotation group $D_{4h}$, either $s$ or $d_{x^2-y^2}$.

Using a Ginzburg-Landau model in which both $s$ and $d_{x^2-y^2}$
order parameter symmetries are allowed, but only one is favored in the
bulk, we find that there are two possibilities consistent with both
the CuO$_2$ plane and $c$-axis tunneling experiments.

The first possibility is that the order parameter is $s$-wave in the
bulk and a $d$-wave component is mixed in at faces normal to the
CuO$_2$ planes (Fig. 1a).  This does not require any special choice of
parameters; there is an instability to mixing near these faces.  The
symmetry being tested is rotation in the CuO$_2$ plane and placing an
edge in that plane breaks the symmetry explicitly.  This always causes
mixing.  The amount of mixing depends on the energetics: if the
$d$-wave component is strongly disfavored (as might be expected in a
conventional superconductor), the mixing is small.  If there is a
close competition, the mixing may be large.  In addition, we find that
for this case the mixing can explain the CuO$_2$ plane experiments
only if the order parameter breaks time reversal invariance at the
surface: it must have the form $s+id$ there.

The second possibility is that the order parameter is $d$-wave in the
bulk and a surface state forms which mixes in an $s$ component on the
face perpendicular to $c$-axis (Fig. 1b).  This occurs only under
certain conditions.  The two components must inhibit each other, in
the sense that the presence of one makes the other energetically less
favorable.  In addition, the effect of the $c$-axis boundary must be
such that the magnitude of the $d$-wave component decreases
significantly from its bulk value near the edge.  In that case, the
$s$-wave component is less suppressed near the surface and a localized
region of mixed symmetry can develop \cite{rokhsar}.

The presence of the surface state normal to the $c$-axis is 
sensitive to the boundary conditions.  This may explain the difficulty
in achieving $c$-axis junctions, as well as the variability among
samples of angle-resolved photoemission spectroscopy studies of the
gap magnitude \cite{shen}.  The photoemission studies see the topmost
CuO$_2$ layer.  Variations in surface properties affect the boundary
conditions, which in turn affect whether the order parameter has the
form $d+s$, $d+is$ or pure $d$ at the surface, each of which
has a different momentum dependence.

The $c$-axis surface state may also lead to ``$\pi$-junction''
behavior.  In a SQUID loop between YBCO and a conventional
superconductor, with junctions normal to the $c$-axis, the
configuration with opposite relative phases on the two junctions will
lead to a net phase difference of $\pi$ in the absence of an applied
magnetic field.

For both possibilities discussed above, the starting point is the
Ginzburg-Landau free energy \cite{glreview}
\begin{eqnarray}
\label{generalgl}
F &=& F_s \ +\  F_d \ + \ F_{sd} \ + \ 
{1\over 8\pi}\, \int \! d^3 {\bf r}\, \; {\bf B}^2 \ ,\\[3\jot]
\nonumber
F_i &=& \int \! d^3 {\bf r}\ \; \Big[\; \kappa_i \, |{\bf D} \psi_i|^2
\ +\  a_i(T)\, |\psi_i|^2\  +\  b_i\, |\psi_i|^4 \;\Big] \\
\nonumber
F_{sd} &=& \int \! d^3 {\bf r}\ \; \Big[\; \lambda_1\,  |\psi_s|^2 |\psi_d|^2
\ +\  \lambda_2 \, (\psi^{\ast 2}_s \psi_d^2 + \psi^{2}_s \psi_d^{\ast 2})
 \; \\
\nonumber
&& \hspace{30pt} +\ \big(\, \gamma\, \psi_d^\ast
(D_x^2 - D_y^2) \psi_s\ +\ {\rm c.c.} \, \big) \; \Big] \ .
\end{eqnarray}
The CuO$_2$ planes are in the $x$-$y$ directions, the magnetic
field ${\bf B} = \nabla \times {\bf A}$, the gauge invariant
gradient operator ${\bf D} = {\bf \nabla} - 2ie{\bf A}/c$, and the
index $i$ runs over $s$ and $d$.
We now consider the two cases separately.

{\bf Bulk $s$ case:} We consider a homogeneous system in
the absence of a magnetic field, for which
the gradient terms in $F$ vanish in the bulk.
A purely $s$-wave solution to Eq. (\ref{generalgl})
exists when $a_s<0$ and $a_d + |a_s|\,(\lambda_1 -2\lambda_2)/2b_s>0$.
If $a_d<0$ and $a_s + |a_d|\,(\lambda_1 -2\lambda_2)/2b_d>0$ then a
purely $d$-wave solution also exists.  As long as $a_s^2/b_s>
a_d^2/b_d$, the $s$ solution has lower energy than the $d$ solution
and is the stable global minimum of the free energy.

Ordinarily, near a boundary, a stability criterion follows from
considering the change in energy due to adding a small $\psi_d$ to the
bulk $\psi_s$ solution: 
\begin{equation}
\label{deltaf}
\delta F = \psi_d^* \Big[
-\kappa_d\,\partial_x^2 + a_d + \lambda |\psi_s|^2 \Big] \psi_d +
O(|\psi_d|^4) \ ,
\end{equation}
where
\begin{equation}
\label{lambdadef}
\lambda \equiv \lambda_1 + 2\lambda_2\cos 2\theta_{sd}\ , 
\end{equation}
$\theta_{sd}$ is the relative phase between the $s$ and $d$ order
parameters, $\psi_s(x)$ is the unperturbed solution, and we consider a
boundary along the $x$ direction.  If the operator in brackets has an
eigenstate with negative eigenvalue, then the energy will be lowered
by forming a surface state.  If there are no negative eigenvalues,
the system is stable against the formation of a surface state.

This stability criterion assumes that the gradient terms mixing the
the $s$ and $d$ components in $F_{sd}$ can be neglected.  For a
boundary which is in the $a$-$b$ plane, this will not necessarily be
true.  Near a boundary, the $s$-wave component may decrease, in which
case the gradient terms in $F_{sd}$ contribute.  These add a term
linear in $\psi_d$ to Eq. (\ref{deltaf}).  The effect of a linear term
is that the energy can always be lowered by turning on a small
$\psi_d\ne 0$, that is, there is an instability to mixing.

This instability occurs only for boundaries in the $a$-$b$ plane
because boundaries in this plane explicitly break the rotational
symmetry being tested: they allow a term such as $\psi_s^*
(D_x^2-D_y^2) \psi_d$ to contribute to the free energy.  For a
boundary along the $c$-axis, there is no equivalent linear-order
mixing through gradients.

The form of the magnitudes $S({\bf r})\equiv |\psi_s({\bf r})|$
and $D({\bf r})\equiv |\psi_d({\bf r})|$ near the boundary will depend
in detail on the Ginzburg-Landau parameters and the boundary conditions. 
The relative phase $\theta_{sd}$
between $\psi_s$ and $\psi_d$ is easier to understand.
There are only two terms in the free energy given in
Eq. (\ref{generalgl}) which depend on this phase:
\begin{equation}
\label{phaseterms}
\lambda_2\; S^2 \! D^2  \cos 2\theta_{sd}
\ + \ \gamma \, D(\partial_x^2-\partial_y^2) S   \, \cos\theta_{sd}\ ,
\end{equation}
where we have assumed there are no spontaneous phase gradients (currents) at
the edge.  The second term is minimized for a relative phase
difference between the $s$ and $d$ components of $0$ or $\pi$
depending on the sign of $\gamma$.  However, for a given system, this
term will prefer opposite phases on the faces normal to the $\bf \hat
x$ and $\bf \hat y$ directions: $s+d$ on one face and $s-d$ on the
other \cite{corner}.  Such a configuration is not consistent with the
corner junction experiments because there will be no net phase shift
in a loop formed between two adjacent faces.

The first term in Eq. (\ref{phaseterms}), however, favors
$\theta_{sd}=\pm \pi/2$ when the coefficient $\lambda_2$ is positive,
and $\theta_{sd}=0$ or $\pi$ when $\lambda_2$ is negative.  If
$\theta_{sd}=\pm \pi/2$ at the minimum then the second term does not
contribute and there is no preference from this surface energy for
either sign.  The corner energy is minimized for a uniform phase
around the material, and the result is $s+id$ at every face or $s-id$
at every face.  This solution, which occurs when the parameters are
such that $\lambda_2 < 0$ and the first term dominates over the second
term at the minimum, has the potential to be consistent with the
corner junction experiments.  

Both the single corner junction and tricrystal ring experiments,
however, place strong limits on the amount of $s$ which is present at
the surface \cite{dvhreview,tsuei}.  In order for the bulk $s$
scenario to explain these results, the parameters must be fine tuned
so that the amount of residual $s$-wave order parameter near the
surface is small.  This makes the picture somewhat unlikely, although
not yet ruled out.

{\bf Bulk $d_{x^2-y^2}$ case:} In this case, we assume that the corner
junction experiments are detecting the intrinsic, bulk order parameter
symmetry and that surface states with mixed symmetry form at the faces
perpendicular to the $c$-axis.  

The energy cost of adding a small $\psi_s$ solution to a bulk $\psi_d$
solution is
\begin{equation}
\label{deltafcaxis}
\delta F = \psi_s^* \Big[
-\kappa_s\,\partial_z^2 + a_s + \lambda |\psi_d|^2 \Big] \psi_s +
O(|\psi_s|^4)\ ,
\end{equation}
where $\lambda$ is defined in Eq. (\ref{lambdadef}) and we now
consider a boundary along the $\bf \hat z$-direction.  When the
operator in brackets develops a bound state with a negative energy
satisfying the appropriate boundary conditions, a
surface state will form.

The boundary conditions determine the presence of the surface state in
the following way.  If $\psi_d$ near the surface decreases very little
from its bulk value, than the $\psi_s$ component will be as suppressed
as it is in the bulk and a surface state is unlikely to form.
Conversely, if $\psi_d$ does decrease significantly, a surface state
may be induced.  More explicitly, the boundary condition at an
interface can be written in general \cite{degennes} as
$d\psi(z)/dz|_{z=0} = \psi(0)/L$, where $L$ measures the extent to
which the order parameter is suppressed at the interface.  At a
superconductor-insulator boundary $L\to \infty$, so $\psi(0)$ is close
to the bulk value, whereas at a superconductor-metal boundary $L\to
0$, so that $\psi(0)$ nearly vanishes.  Therefore boundaries
which are more superconductor-metal like enhance the likelihood
of a surface state forming.

The symmetry of the surface state will depend on the sign
of the $\lambda_2$ defined in Eq. (\ref{generalgl}):
the state will have the form $s\pm i d$ for positive $\lambda_2$ 
and $s\pm d$ for negative $\lambda_2$ \cite{rokhsar}.
For a given system (fixed $\lambda_2$), the plus 
and minus states are degenerate, and this can lead to the effects 
illustrated in Figs. \ref{figopsquid} and \ref{figopfofphi}.
The Josephson energy for the SQUID loops shown
in Fig. \ref{figopsquid} can be written
\begin{equation}
E_J(\Phi) = E_{J,a}\;\cos\phi_a \ + \ \eta\ E_{J,b}\cos\phi_b\ ,
\end{equation}
where $E_{J,i}$ are the Josephson coupling energies
for the top and bottom junctions ($i=a, b$),
$\phi_i$ are the phase differences across these junctions, and
$\eta$ is $\pm 1$.
The phase differences satisfy
\begin{equation}
\phi_a-\phi_b \ =\  2\pi\; {\Phi\over \Phi_0}\ ,
\end{equation}
where $\Phi$ is the net flux enclosed in the loop and $\Phi_0=hc/2e$
is the flux quantum \cite{tinkham}.  The configuration in which the
top and bottom junctions have misaligned $s$ components ($\eta=-1$)
adds an additional phase shift of $\pi$ because the conventional
superconductor couples only to the $s$ component of the order
parameter (assuming a tunnel junction in which higher order tunneling
matrix elements can be neglected).  The result, for symmetric
junctions, is shown in Fig. \ref{figopfofphi}.  Minimizing the
Josephson energy yields a net phase shift of $\Phi_0/2$ between the flux
dependence of the $\eta=+1$ and $\eta=-1$ configurations. 

This phase shift leads to several interesting behaviors. 
In the high-inductance limit, the $\eta=-1$ configuration in Fig.\
\ref{figopsquid} will cause spontaneous
currents generating half-integral flux quantua,
as in the tricrystal ring configurations.  The half-integral
periodicity of $E_J^{}(\Phi)$ is also reflected in a half-integral
periodicity in the flux dependence of the critical current, which will
be seen if the system can probe both configurations in achieving
the maximum supercurrent.  We also note that an asymmetric junction,
for which $E_1 \ne E_2$, causes the resulting $E_J^{}(\Phi)$ to
oscillate between $-|E_1-E_2|$ and $-|E_1+E_2|$, instead of $0$ and
$-2E_1$ as in Fig. \ref{figopfofphi}, but the periodicity with
$\Phi_0/2$ remains the same.

There are several other interesting consequences of a mixed symmetry
surface state.  First, for the state normal to the $c$-axis, there
should be a surface phase transition, at a temperature $T_{c,s}$ below
the bulk $T_c$, where the $s$-wave component of the order parameter
disappears.  This would most likely occur at too high a temperature to
be seen in the Josephson tunneling into a conventional superconductor,
but might be seen in photoemission.  There is some preliminary
evidence from photoemission studies on $\rm Bi_2 Sr_2 Ca Cu_2 O_{8+x}$
that at a temperature $\approx .8$ to $.9\, T_c$ there is an increase
in anisotropy of the gap magnitude, as would occur if a surface
$s$-wave contribution to the order parameter were disappearing
\cite{ma}.  Second, the relative phase angle between the two
components, $\theta_{sd}$, is a dynamical variable and its
oscillations, which occur as long as there is a charging energy (a
capicitance in the Josephson equations), are a new collective mode of
the order parameter.  This mode is charged, since $|\psi_s|\ne
|\psi_d|$, and will therefore be pushed up near the plasma frequency,
but is distinct from the usual plasma mode.  The mode will disappear
at the temperature $T_{c,s}$ of the surface phase transition.

This work was stimulated by many interesting discussions with
S.~Kivelson and V.~Emery.  I would also like to thank D.~Rokhsar for
valuable discussions and J.~Clarke, J.-W.~Gan, and D.-H.~Lee for
helpful comments.  This work was supported by a Miller Research
Fellowship from the Miller Institute for Basic Research in Science.

\newpage

\bigskip
\begin{figure}
\epsfysize=7.5cm\epsfxsize=8cm\centerline{\epsfbox{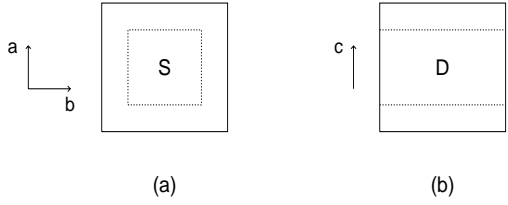}}
\caption{Two possiblities consistent with both sets of phase sensitive
experiments: 
(a) A surface state of mixed $s$ and $d$ symmetry forms 
normal to the CuO$_2$ planes in a bulk $s$-wave superconductor;
(b) A surface state of mixed symmetry forms 
parallel to the CuO$_2$ planes in a bulk $d$-wave superconductor.}
\label{figoptwo}
\end{figure}

\bigskip
\begin{figure}
\epsfysize=8cm\epsfxsize=9cm\centerline{\epsfbox{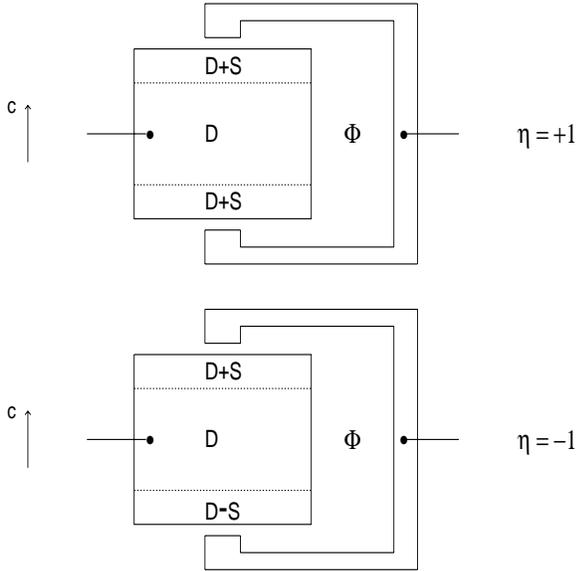}}
\caption{Two possible configurations of a $d$-wave superconductor, with
surface states of mixed symmetry, in a SQUID loop with a conventional
superconductor.}
\label{figopsquid}
\end{figure}

\bigskip
\begin{figure}
\epsfysize=5cm\epsfxsize=7cm\epsfbox{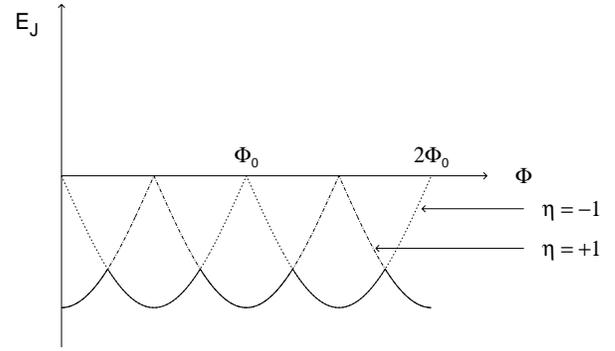}
\caption{Josephson energy of the two configurations in Fig. 
\protect{\ref{figopsquid}}.  The $\eta=+1$ configuration behaves
as an ordinary SQUID, but the flux dependence for the 
$\eta=-1$ configuration is shifted by $\Phi_0/2$.  The result is
that the ground state energy of the system (solid line) 
is periodic with period $\Phi_0/2$ rather than $\Phi_0$.  }
\label{figopfofphi}
\end{figure}



\begin{references}
\bibitem{dvhreview} Reviewed in D.~J.~Van~ Harlingen, Rev. Mod.
Phys. {\bf 67}, 515 (1995).
\bibitem{wollman} D.~A.~Wollman {\it et al.}, Phys. Rev. Lett. {\bf
71}, 2134 (1993)
\bibitem{brawner} D.~A.~Brawner and H.~R.~Ott, Phys. Rev. B {\bf 50},
6530 (1994).
\bibitem{mathai} A.~Mathai {\it et al.}, Phys. Rev. Lett. {\bf 74},
4523 (1995).
\bibitem{wollmanfm} D.~A.~Wollman {\it et al.}, Phys. Rev. Lett. 
{\bf 74}, 797 (1995).
\bibitem{miller} J.~H.~Miller {\it et al.}, Phys. Rev. Lett. {\bf 74},
2347 (1995).
\bibitem{tsuei} C.~C.~Tsuei {\it et al.}, Phys. Rev. Lett. {\bf 73},
593 (1994); J.~R.~Kirtley {\it et al.}, Nature {\bf 373}, 225 (1995).
\bibitem{dynes} A.~G.~Sun, D.~A.~Gajewski, M.~B.~Maple, and R.~C.~Dynes,
Phys. Rev. Lett. {\bf 72}, 2267 (1994).
\bibitem{iguchi} I.~Iguchi and Z.~Wen, Phys. Rev. B {\bf 49}, 12388 (1994).
\bibitem{clarke} 
R. Kleiner {\it et al.}, Phys. Rev. Lett., to appear.
\bibitem{dontconsider}
We do not consider here the abundant evidence for nodes in the gap function,
which is consistent with either symmetry when variations with $|{\bf k}|$ are
allowed.
\bibitem{scalapino} 
For a review of the relevance of
the order parameter symmetry to the microscopic pairing mechanism, see
D.~J.~Scalapino, Phys. Rep. {\bf 250}, 329 (1995).
\bibitem{rokhsar}
See also D.~S.~Rokhsar, preprint.
\bibitem{shen}
Reviewed in Z.~X.~Shen and D.~S.~Dessau, Phys. Rep. {\bf 253}, 1 (1995).
(See Fig. 5.18.)
\bibitem{glreview} 
For recent reviews of the
Ginzburg-Landau description of order parameter mixing 
in superconductors, see L.~P.~Gor'kov, 
Sov. Sci. Rev. A {\bf 9}, 1 (1987) and 
M.~Sigrist and K.~Ueda, Rev.\ Mod.\ Phys.\ {\bf 63},
239 (1991).  See also M.~Sigrist, D.~B.~Bailey, and R.~B.~Laughlin,
Phys. Rev. Lett. {\bf 74}, 3249 (1995).
\bibitem{corner} There will be an energy cost at the corner where these
two configurations intersect, but we assume that this corner energy
is much less than the surface energy represented in Eq.
(\ref{phaseterms}).
\bibitem{degennes} See, {\it e.g.}, P.~G.~de~Gennes, 
 {\it Superconductivity of Metals and Alloys}, Benjamin, New York, 1966.
\bibitem{tinkham} See, {\it e.g.}, M.~Tinkham, {\it Introduction to
Superconductivity}, McGraw-Hill, New York, 1975.
\bibitem{ma}
J.~Ma {\it et al.}, Science {\bf 267}, 862 (1995).
\end{references}
\end{document}